\documentclass[12pt,a4paper]{article}
\usepackage{epsfig}
\usepackage{units}

\begin{document}

\newcommand{\jpsi}{J/\psi}
\newcommand{\chicj}{\chi_{cJ}}
\newcommand{\chiczero}{\chi_{c0}}
\newcommand{\chicone}{\chi_{c1}}
\newcommand{\chictwo}{\chi_{c2}}
\newcommand{\psip}{\psi^{\prime}}
\newcommand{\psipp}{\psi^{\prime\prime}}
\newcommand{\gm}{\gamma}
\newcommand{\gmsm}{\gamma_{\unit{sm}}}
\newcommand{\gmlg}{\gamma_{\unit{lg}}}
\newcommand{\dil}{\ell^+\ell^-}
\newcommand{\ee}{e^+e^-}
\newcommand{\mm}{\mu^+\mu^-}
\newcommand{\mev}{\,\unit{MeV}}
\newcommand{\mevcc}{\,\unit{MeV}/\unit{c}^2}
\newcommand{\gev}{\,\unit{GeV}}
\newcommand{\gevc}{\,\unit{GeV}/\unit{c}}
\newcommand{\gevcc}{\,\unit{GeV}/\unit{c}^2}
\newcommand{\br}[1]{\mathcal{B}(#1)}
\newcommand{\brb}{\mathcal{B}}
\newcommand{\chisq}[1]{\chi^{2}_{\unit{#1}}}
\newcommand {\eg}       {\emph{e}.\emph{g}.}
\newcommand {\ie}       {\emph{i}.\emph{e}.}
\newcommand {\cf}       {\emph{cf}.}

\title{\huge \bf Recent Progress on Charmonium Decays at BESIII \footnote{\footnotesize Proceedings of: 5$^{{\mathrm t}{\mathrm h}}$ International Conference in High-Energy Physics: HEP-MAD 11,
25-31 August 2011, Antananarivo, Madagascar}}

\author{{\Large Xiao-Rui Lu} \\
{\normalsize ({\it on behalf of the BESIII Collaboration})}\\
Physics Department\\
Graduate University of Chinese Academy of Sciences\\
Beijing, 100049, China\\
{\normalsize \it xiaorui@gucas.ac.cn}}
\date{}
\maketitle

\begin{abstract}

In 2009, the BESIII experiment has collected about 225M $\jpsi$ and 106M $\psip$ samples, both of which are the world largest on-peak charmonium production. Based on these dataset, BESIII has made great effort on the study of the charmonium decays, some important of which have been reviewed in this proceeding. In addition, a searching for new physics through the $CP/P$ violation process is reported.

\end{abstract}

\section{Introduction}

BEPCII/BESIII is an upgrade facility from the previous
BEPC/BES~\cite{:2009vd}. The collider experiment BESIII is monitoring the double-ring electron-positron head-on collisions produced by the BEPCII machine. The luminosity is optimized at a center-of-mass energy of 3.78\gev and the peak record has reached to $6.4\times 10^{32}$cm$^{-2}$s$^{-1}$, which luminosity is one order of
higher than that at CESR-c~\cite{Kubota:1991ww}. The beam energy ranges from 1.0$\gev$ to 2.3$\gev$. Therefore, the physics in BESIII cover the
$\tau$-charm physics.

\begin{figure}[htbp]
\centerline{\includegraphics[angle=90,width=0.9\textwidth]{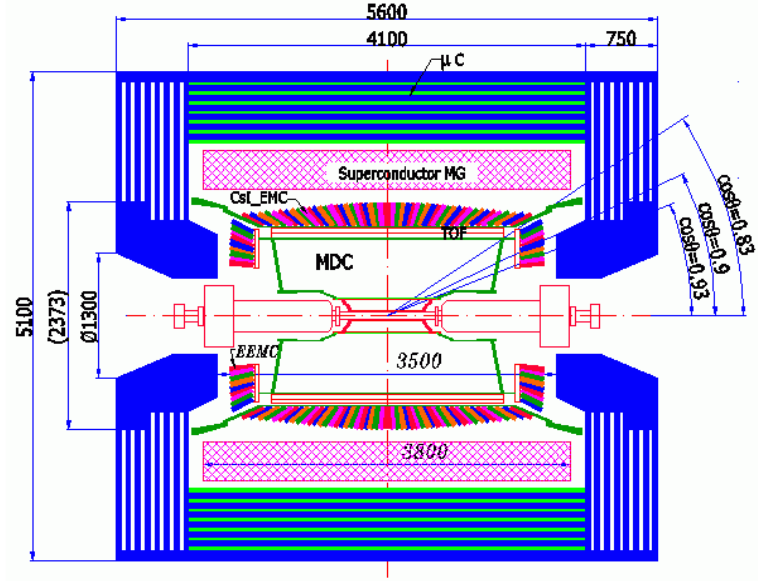}}
\vspace{-1cm}
\caption{An Overview of the BESIII Detector.}
\label{fig:bes}
\end{figure}

The BESIII spectrometer~\cite{:2009vd} as shown in Fig.~\ref{fig:bes}, consists of the following main components with order of the distance to the interaction position: a main draft chamber with momentum resolution 0.5\% at 1\gev; an electromagnetic calorimeter
with energy resolution 2.5\% at 1.0\gev; a Time-Of-Flight counters; a superconducting magnet with a field of 1\,T; a muon chamber system made of resistive plate chambers.

In 2009, the BESIII experiment has collected about 225M $\jpsi$ and 106M $\psip$ samples, both of which are the world largest on-peak charmonium production. Based on these dataset, BESIII has made great effort on the study of the charmonium decays, some important of which have been reviewed in this proceeding. In addition, a search for new physics through the $CP/P$ violation process is reported.

\section{Study of Radiative Decays $\psip\to\gm+P$($P=\pi^0$, $\eta$, $\eta^\prime$)}

The radiative decay of the $\psip$ to a pseudo-scalar meson provide important tests for various mechanisms in understanding the low-$Q^2$ phenomena, such as the vector meson dominance model (VDM), the two-gluon coupling to $c\bar c$ states, the $\eta - \eta^\prime$ mixing angle, the final-state radiation by light quarks. Recently, the
CLEO-c Collaboration reported measurements for the decays of $\jpsi$, $\psip$, and $\psipp$ to $\gm P$~\cite{:2009tia}, and no evidence for $\psip\to\gm\eta$ or $\gm\pi^0$ was found.

\begin{figure}[hbtp]

\begin{minipage}[t]{0.45\linewidth}
    \centering
    \includegraphics[width=0.94\linewidth, height=0.7\textwidth]{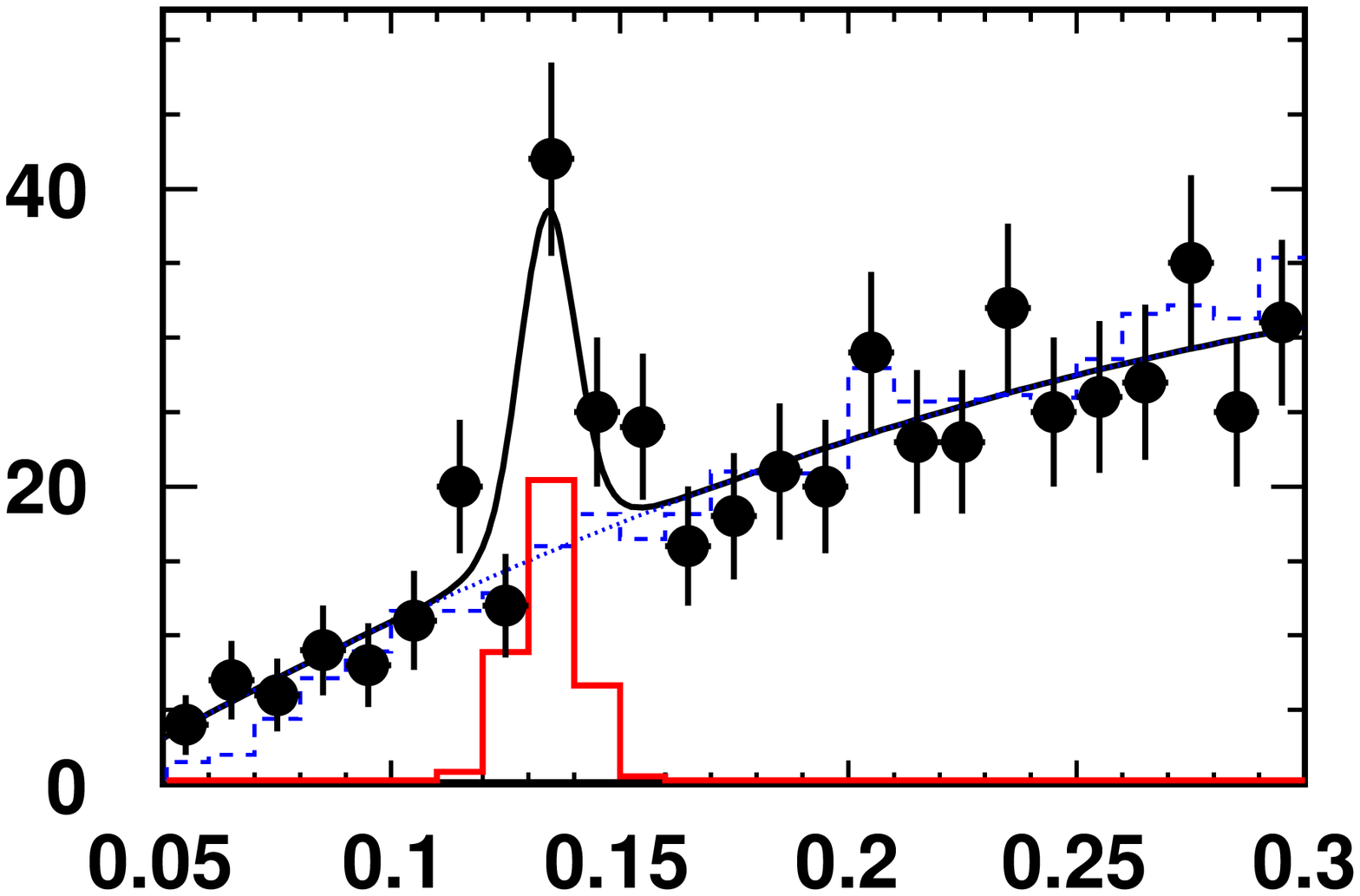}
\end{minipage}
\put(-160, 105){ \bf (a)}

\begin{minipage}[t]{0.45\linewidth}
    \centering
    \ \
\end{minipage}

\begin{minipage}[t]{0.45\linewidth}
   \centering
    \includegraphics[width=0.95\linewidth, height=1.4\textwidth]{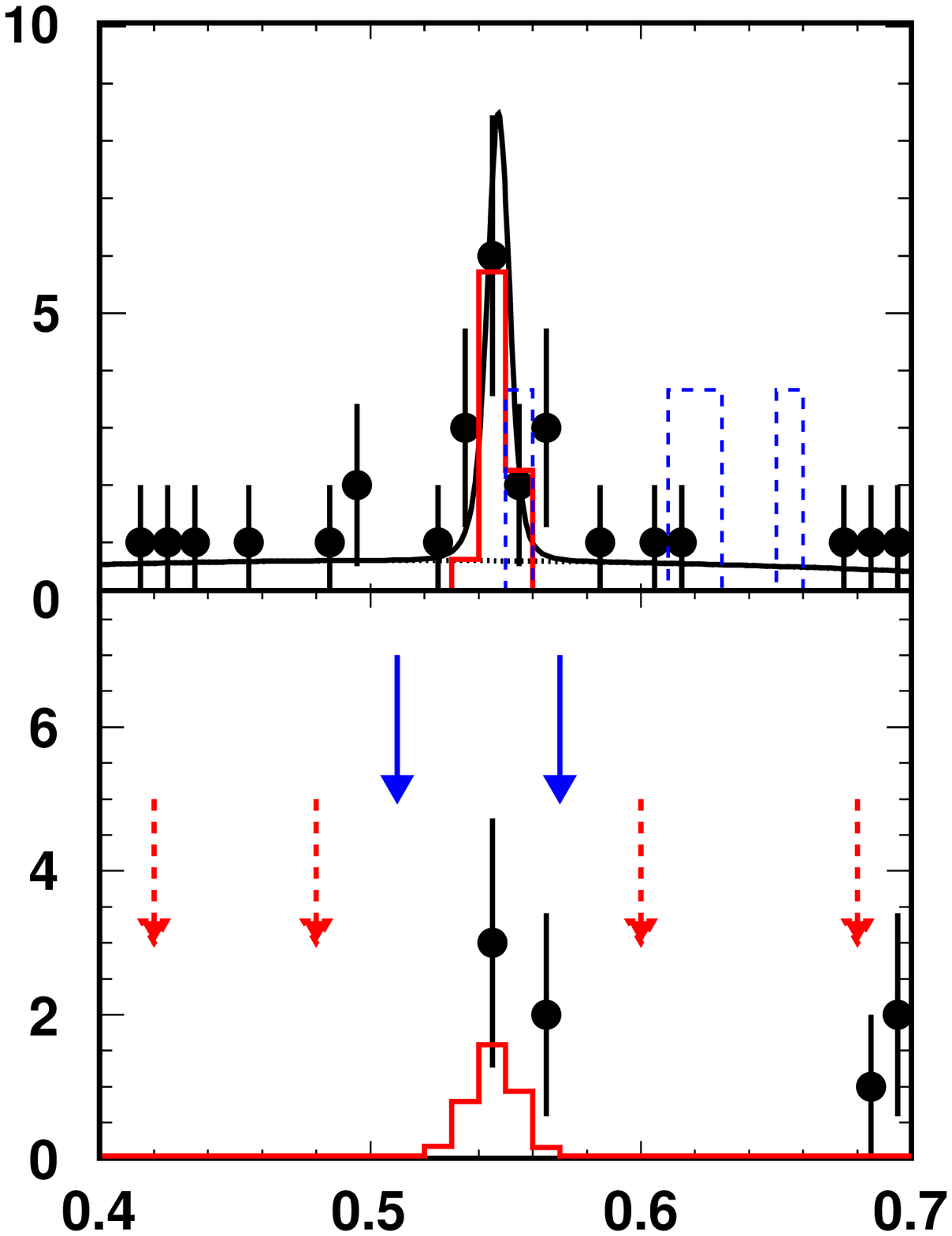}
\end{minipage}
\hspace{-0.4cm}
\begin{minipage}[t]{0.45\linewidth}
    \centering
    \includegraphics[width=0.95\linewidth, height=1.4\textwidth]{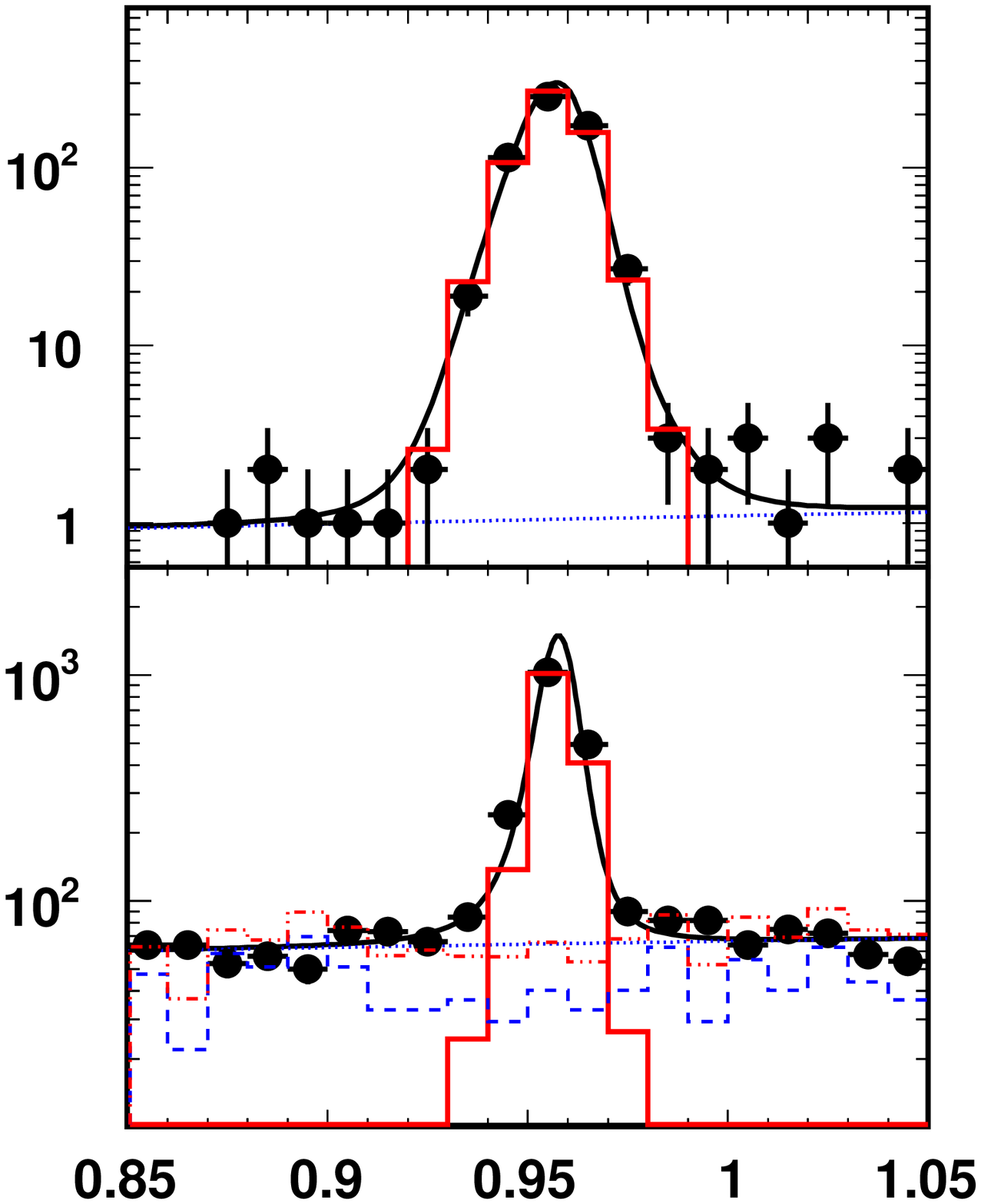}
    \put(-350,250){ \bf (b)}
    \put(-350,150){ \bf (c)}
    \put(-155,250){ \bf (d)}
    \put(-155,150){ \bf (e)}
    \put(-235,39){ \textbf{Mass (\gev)}}
    \put(-380,110){\rotatebox{90}{ \textbf{Events/(10 \mev)}}}
    \end{minipage}
\vspace{-1.5cm}
    \caption{Mass distributions of the pseudoscalar meson candidates for $\psi^\prime\rightarrow \gamma P$:
a) $\gm\pi^0$; b) $\gm\eta(\pi^+\pi^-\pi^0)$; c) $\gm\eta(
3\pi^0)$; d) $\gm\eta^\prime(\pi^+\pi^-\eta(\gm\gm))$; e)
$\gm\eta^\prime(\gm\pi^+\pi^-)$. Refer to Ref.~\cite{Ablikim:2010dx} for details.}
    \label{fig:psip-gp}
\end{figure}

BESIII studied the processes
$\psip\to\gm\pi^0$ with $\pi^0\to\gm\gm$, and $\psip\to\gm\eta$ with $\eta\to\pi^+\pi^-\pi^0[\pi^0\pi^0\pi^0]$, and $\psip\to\gm\eta^\prime$ with $\eta^\prime\to\gm\pi^+\pi^-$. The analyses use the 106M $\psip$ data sample. The results show that $\psip\to\gm\pi^0$ and $\psip\to\gm\eta$ are observed for the first time with
significance of 4.6$\sigma$ and 4.3$\sigma$, respectively, and with
branching fractions of
$\br{\psip\to\gm\pi^0}=(1.58\pm0.40\pm0.13)\times
10^{-6}$ and $\br{\psip\to\gm\eta}=(1.38\pm0.48\pm0.09)\times 10^{-6}$. Branching fraction of the process $\psip\to\gm\eta^\prime$
is measured with improved accuracy to be $\br{\psip\to\gm\eta^\prime}=(126\pm3\pm8)\times10^{-6}$. The mass distributions of the pseudoscalar meson
candidates are shown in Fig.~\ref{fig:psip-gp}. For the first time, BESIII determined the ratio of the $\eta$ and $\eta^\prime$ production rates from $\psip$ decays,
$R_{\psip}\equiv B(\psip\to\gm\eta)/B(\psip\to\gm\eta^\prime)=(1.10\pm0.38\pm0.07)$\%. This ratio is in agreement with
90\% C.L. upper limit determined by CLEO-c and in
contradiction to predictions of leading-order perturbative QCD. This is also smaller than the corresponding ratio for the
$J/\psi$ decays by an order of magnitude, which was measured to be $R_{J/\psi}=(21.1\pm0.9)$\%. More details on this study can be found in Ref.~\cite{Ablikim:2010dx}.

\section{Study of Hadronic Decays of $\chicj$ }

BESIII has world largest $\psip$ on-peak production, which endow BESIII with the largest radiative decay $\chicj$ data. $\chicj$ hadronic decay helps in understanding the $P$-wave charmonium decay dynamics. Most of hadronic decay channels of $\chicj$ are not still known and the pursued measurement provide a good laboratory to test the color singlet and octet mechanism in interpreting $\chicj$ hadronic decays. At BESIII, $\chicj$ decays to $PP(P=\pi^0, \eta)$~\cite{Ablikim:2010zn}, $VV(V=\omega, \phi)$~\cite{:2011ih}, $\Lambda\bar{\Lambda}$~\cite{Collaboration:2011uf} and $4\pi^0$~\cite{Ablikim:2010jr} have been studied through $\psip\to\gm\chicj$ decays.

The decay $\chicj$ decays to $VV(V=\omega, \phi)$ were studied
with $\phi$ reconstructed form $K^+K^-$ or $\pi^+\pi^-\pi^0$, $\omega$ from $\pi^+\pi^-\pi^0$, and $\pi^0$ from $\gm\gm$. The results are shown in Table~\ref{tab:chicjvvres}. The decays of $\chicone\to\phi\phi[\omega\omega]$, which were suppressed in helicity sum rule (HSR)~\cite{Brodsky:1981kj}, and the doubly OZI suppressed decay $\chiczero\to\omega\phi$ are observed for the first time. Evidence for the $\chicone\to\omega\phi$ decay is found with a signal significance of $4.1\sigma$. The branching fractions for $\chi_{c0[2]}\to\phi\phi[\omega\omega]$ decays are remeasured with improved precision. These precise measurements will be helpful for understanding $\chicj$ decay mechanisms. In particular, the measured branching fractions for $\chi_{c1}\to VV$ indicate that HSR is significantly violated and that long distance effects play an important role in this energy region. Details about this study can be found in Ref.~\cite{:2011ih}.

\begin{table}[htbp]
\caption{Summary of the branching fractions ($\mathcal B$)
for $\chicj\to VV$ measured at BESIII~\cite{:2011ih}. Also listed are the world average in PDG~\cite{Nakamura:2010zzi}. The upper limit is estimated
at the 90\% C.L.\label{tab:chicjvvres}}
\begin{center}
\begin{tabular}{l|l|l}
 \hline\hline
 Mode& PDG $\mathcal B(\times10^{-4}$)& BESIII $\mathcal B(\times10^{-4}$) \\\hline
$\chiczero\to\phi\phi$&$9.2\pm1.9$&$8.0\pm0.3\pm0.8$\\
$\chicone\to\phi\phi$&&$4.4\pm0.3\pm0.5$\\
$\chictwo\to\phi\phi$&$14.8\pm2.8$&$10.7\pm0.3\pm1.2$\\\hline
$\chiczero\to\omega\omega$&$22\pm7$&$9.5\pm0.3\pm1.1$\\
$\chicone\to\omega\omega$&&$6.0\pm0.3\pm0.7$\\
$\chictwo\to\omega\omega$&$19\pm6$&$8.9\pm0.3\pm1.1$\\ \hline
$\chiczero\to\omega\phi$&&$1.2\pm0.1\pm0.2$\\
$\chicone\to\omega\phi$&&$0.22\pm0.06\pm0.02$\\
$\chictwo\to\omega\phi$&&$<0.2$ \\
\hline
 \hline
\end{tabular}
\end{center}
\end{table}

Hadronic decay of $\chicj$ to baryon pair has challenged the color octet mechanism, since many of baryon pair decays present disagreement between experimental measurements with theoretical calculations, in particular $\chicj\to\Lambda\bar{\Lambda}$. BESIII studied the process $\chicj\to\Lambda(1520)\bar{\Lambda}(1520)$ decaying to $p\bar{p}K^{+}K^{-}$ with results of
$\br{\chiczero\to \Lambda(1520)\bar{\Lambda}(1520)}=(3.18\pm1.11\pm0.53)\times 10^{-4}$,
$\br{\chicone\to \Lambda(1520)\bar{\Lambda}(1520)}<1.00\times 10^{-4}$ and $\br{\chictwo\to \Lambda(1520)\bar{\Lambda}(1520)}=(5.05\pm1.29\pm0.93)\times 10^{-4}$. This is the first measurement on the $\chicj$ decays to excited baryon pairs. Details can be found in Ref.~\cite{Collaboration:2011uf}.

\section{Study of Radiative Decays $\chicj\to\gm V(V=\rho, \omega, \phi)$ }

Doubly radiative decays of the type $\psi\to\gamma X\to \gamma\gamma V$, provide information on the flavor content of the $C$-even resonance $X$ and
on the gluon hadronization dynamics in the
process~\cite{Ablikim:2011kv}. For the case $X = \chicj$, it may provide an independent
window for understanding possible glueball dynamics and validating
theoretical techniques. The vector meson of $\rho$, $\omega$ and $\phi$ were studied at BESIII.

Table~\ref{tab:chicjrVres} shows the numerical results, along with the theoretical predictions from perturbative quantum chromodynamics (pQCD)~\cite{zhaogd}, nonrelativistic QCD
(NRQCD)~\cite{zhaogd1}, and NRQCD plus QED contributions
(NRQCD+QED)~\cite{zhaogd1}, and the results from the CLEO
experiment~\cite{Bennett:2008aj}. We see the improved measurements at BESIII provide tighter constraints on theoretical
calculations. Especially, we observe $\chi_{c1}\to \gamma\phi$ for the first time. In addition, the fraction of the transverse
polarization component of the vector meson in $\chi_{c1}\to \gamma
V$ decay is studied, which indicates the dominant longitudinal component. Details of this study can be found in Ref.~\cite{Ablikim:2011kv}.

\begin{table}[hbtp]
\caption{\label{tab:chicjrVres} Measurements of $\br{\chicj\to VV }$ (in units of $10^{-6}$) at BESIII, and comparison of theoretical predictions   and measurements at CLEO-c. The upper limits are at the 90\% confidence level (C.L.).}
\begin{tabular}{ccccccc} \hline \hline
    Mode & BESIII & CLEO~\cite{Bennett:2008aj}   & pQCD~\cite{zhaogd}  & NRQCD~\cite{zhaogd1}
            & NRQCD\\
         &&&&&+QED~\cite{zhaogd1}\\ \hline
 $\chi_{c0}\to\gamma\rho^{0}$ & $<$16.2 & $<9.6$               & 1.2 &3.2 &2.0 \\
 $\chi_{c1}\to\gamma\rho^{0}$  &$25.8\pm5.2\pm2.3$ & $243\pm 19\pm 22$    & 14  &41  &42  \\
 $\chi_{c2}\to\gamma\rho^{0}$ &$<$8.1 & $<50$                & 4.4 &13  &38  \\
 \hline
 $\chi_{c0}\to\gamma\omega$ & $<$10.5 &  $<8.8$             & 0.13 &0.35&0.22 \\
 $\chi_{c1}\to\gamma\omega$ & $228\pm13 \pm22$ & $83\pm 15\pm 12$    & 1.6  &4.6 &4.7  \\
 $\chi_{c2}\to\gamma\omega$ & $<$20.8 &  $<7.0$             & 0.5  &1.5 &4.2  \\
 \hline
 $\chi_{c0}\to\gamma\phi$ &  $<$12.9 &  $<6.4$               & 0.46&1.3&0.03   \\
 $\chi_{c1}\to\gamma\phi$ &$69.7\pm7.2\pm6.6$ &  $<26$                & 3.6&11&11       \\
 $\chi_{c2}\to\gamma\phi$ &$<$6.1 &  $<13$                & 1.1&3.3&6.5     \\ \hline \hline
\end{tabular}
\end{table}

\section{Study of $\eta^\prime_c$ Decays into Two Vector Mesons}

$\eta^\prime_c$ is only recently established in charmonium family and it's property is intriguing both in experiment and theory. The sizes of the decays of $\eta^\prime_c\to VV$, where $V$ stands for light vector
mesons, are predicted to be highly suppressed under the mechanism of helicity selection rule (HSR). However, the effect from charmed meson loop mechanism can evade HSR effect, and enhance the production. Hence, the measurement of
$\br{\eta^\prime_c\to VV}$ may help in understanding the role
played by charmed meson loops in $\eta_c\to VV$.

BESIII searched for the $\eta^\prime_c$ signls in three exclusive decay channels:
$\psip\to \gm\rho^{0}\rho^{0}\to \gm2(\pi^+\pi^-)$, $\psip\to \gm
K^{*0}\bar{K}^{*0}\to \gm\pi^+\pi^-K^+K^-$, and $\psip\to \gm\phi\phi\to \gm2(K^+K^-)$~\cite{Collaboration:2011kr}. As a result, no obvious $\eta^\prime_c$ signal was observed. The upper limits are given as $\br{\eta^\prime_c\to \rho^{0}\rho^{0}}<3.1\times10^{-3}$,
$\br{\eta^\prime_c\to K^{*0}\bar{K}^{*0}}<5.4\times10^{-3}$, and
$\br{\eta^\prime_c\to \phi\phi}<2.0\times10^{-3}$. These upper limits are lower than the theoretical
predictions. Details of this study can be found in Ref.~\cite{Collaboration:2011kr}.

\section{Study of $CP$ and $P$ Violation through $P\to\pi\pi(P=\eta, \eta^\prime, \eta_c)$}

In Standard Model (SM), $CP and P$ violating process $P\to\pi\pi$ ($P$ is a pseudoscalar meson) can proceed only via the weak interaction with a branching fraction of order $10^{-27}$. Improved QCD only allow up to $10^{-15}$. The decay rates of $P$ and $CP$ invariance process can be experimentally tested. Any higher level will trigger new physics beyond SM.

With world largest $\jpsi$ on-peak data, BESIII searched for signals of the $CP/P$ violating process $P\to\pi^+\pi^-[\pi^0\pi^0](P=\eta, \eta^\prime, \eta_c)$, via $\jpsi\to\gm P$ decays. Table~\ref{tab:P2PiPires} lists the results for the upper limits on
the branching fractions of all the processes studied. This measurement improves the world-best upper limits of the studied channels, except for $\br{\eta \to \pi\pi}$.
These results provide
experimental limits for theoretical models predicting how much $CP$ and
$P$ violation there may be in $\eta^\prime$ and $\eta_c$ meson decays.
Details of this search are described in Ref.~\cite{Ablikim:2011vg}.

\begin{table}[htbp]
\caption{Summary of the limits on $\eta/\eta^\prime/\eta_c$ decays to $\pi^0\pi^0$
and $\pi^0 \pi^0$ states. $\brb^{\rm UP}$ is the upper limit at the 90\% C.L. on
the decay branching fraction of $\eta/\eta^\prime/\eta_c$ to $\pi^+\pi^-$ or $\pi^0\pi^0$, and $\brb^{\rm UP}_{PDG}$ is the upper limit on the decay
branching fraction from PDG~\cite{Nakamura:2010zzi}.} \label{tab:P2PiPires}
\begin{center}
\begin{tabular}{c   c  c  }
\hline
 mode  & $\brb^{\rm UP}$   &$\brb^{\rm UP}_{PDG}$ \\
\hline$\eta \to\pi^+\pi^-$  & $3.9\times 10^{-4}$  &$1.3\times 10^{-5}$  \\
$\eta^\prime \to \pi^+\pi^-$  & $5.5\times 10^{-5}$ & $2.9 \times 10^{-3}$ \\
$\eta_c \to \pi^+\pi^-$ & $1.3 \times 10^{-4}$ &$6 \times 10^{-4}$\\
$\eta \to \pi^0 \pi^0$ & $6.9\times 10^{-4}$  &$3.5\times 10^{-4}$  \\
$\eta^\prime \to \pi^0 \pi^0$ & $4.5\times 10^{-4}$ & $9 \times 10^{-4}$ \\
$\eta_c \to \pi^0 \pi^0$ &  $4.2\times 10^{-5}$  &$4\times 10^{-4}$\\
 \hline
\end{tabular}
\end{center}
\end{table}

\section{Summary}

We report the recent progress on the charmonium decays studied at BESIII, including $\psip$ radiative decays to a pseudoscalar meson, $\chicj$ hadronic decays, $\chicj$ radiative decays to a vector meson, and $\eta^\prime_c$ decays to vector meson pairs. In addition, the search effort for new physics through $CP/P$ violation process, pseudoscalar meson decay to $\pi\pi$, is presented.

\section{Acknowledgements}
This work is supported in part by the National Natural Science Foundation of China (10905091), SRF for ROCS of SEM,
and China Postdoctoral Science Foundation.

\end{document}